\documentclass[dvips,reprint,a4paper,superscriptaddress,twocolumn,prb,showpacs]{revtex4}

\usepackage{amsmath}    
\usepackage{verbatim}   
\usepackage{color}      
\usepackage{subfigure}  
\usepackage{hyperref}   

\begin{document}

\title{Hydrodynamic theory of surface excitations of three-dimensional topological insulators}
\author{N.M. Vildanov}
\affiliation{I.E.Tamm Department of Theoretical Physics,
P.N.Lebedev Physics Institute, 119991 Moscow, Russia}


\begin{abstract}
Edge excitations of a fractional quantum Hall system can be
derived as surface excitations of an incompressible quantum
droplet using one dimensional chiral bosonization. Here we show
that an analogous approach can be developed to characterize
surface states of three-dimensional time reversal invariant
topological insulators. The key ingredient of our theory is the
Luther's multidimensional bosonization construction.
\end{abstract}

\maketitle

Topological insulators are materials with insulating bulk and
topologically protected metallic edge or surface
states.\cite{review1,review2,qi} Two-dimensional TIs are also
called as quantum spin Hall systems. Edge states of QSH systems
are analogous to quantum Hall edge states. However there is an
important distinction\cite{qi,edge_states}: edge states of the QHE
with up and down spins propagate in the same direction, whereas
QSH edge states with opposite directions of spin
counter-propagate. This is the consequence of the time reversal
invariance of QSHE, which is broken in conventional QH systems.
Therefore the former are called chiral and the latter are called
helical edge states.

There are two classes of non-trivial time reversal invariant TIs
in three-dimensions which are called strong and
weak.\cite{fkm,moore,roy} While weak TIs are layered 2D QSH states
(in the sense that these two systems can be adiabatically coupled
to each other), strong TIs are purely three-dimensional. Surface
states of these TIs are massless 2D Dirac fermions. Weak and
strong TIs are distinguished by the number of Dirac cones on the
surface: strong TIs have an odd number and weak TIs have an even
number of Dirac cones on the surface. Gapless modes of strong TIs
are robust and insensitive to weak interactions and disorder.

Surface states of three-dimensional time reversal invariant TIs
are spin filtered, such that
$\langle\vec{s}(-\mathbf{k})\rangle=-\langle\vec{s}(\mathbf{k})\rangle$,
which means that spin density and charge current are coupled (see,
e.g., Ref.[\onlinecite{collective_modes}]). This suggests that
three dimensional strong TIs realize quantum spin Hall effect in
every radial direction. ``Radial direction" here means that one
considers tomographic projection of the surface states on certain
direction. This tomographically projected state in certain cases
can be viewed as a two dimensional quantum spin Hall system. In
this paper we explicitly show how to construct such mapping
mathematically. In constructing this mapping we assume that a
hydrodynamic theory of the QSHE edges can be developed in the
similar way as for QHE edges.\cite{haldane,lee&wen,stone} Several
interesting physical quantities characterizing three-dimensional
TIs are introduced.

It should be made clear that many of the ideas presented here
taken separately are not new. We have only combined them in a
single picture. We also note that this paper is illustrative
rather than rigorous.

A good description of the hydrodynamic theory of the FQHE edge
states can be found in [\onlinecite{wen}]. Here we briefly mention
the main points. Then we modify this theory to describe surface
states of strong topological insulators.

Suppose that the FQHE states are incompressible irrotational
liquid without bulk excitations. Then the only low lying
excitations are the surface deformations of the quantum droplet.
The droplet is confined by a smooth potential well. The electric
field of the potential well will generate a persistent current
along the edge fluctuating part of which is given by
$$
\mathbf{j}=\sigma_{xy}[\mathbf{e}_z,\mathbf{E}]h(x),\qquad
\sigma_{xy}=\nu\frac{e^2}{h}
$$
where $\nu$ is the filling fraction, $\mathbf{e}_z$ is the unit
vector along the z axis, $h(x)$ is the displacement of the edge
from its equilibrium value, $x$ is the coordinate along the edge.

One-dimensional density of the edge wave $\rho(x)$ is related to
the displacement of the edge $h(x)$ through $\rho(x)=nh(x)$, where
$n=\nu\frac{eH}{hc}$ is the two dimensional electron density in
the bulk. Then continuity equation reads

\begin{equation}\label{continuity}
    \partial_t h-v\partial_x h=0
\end{equation}
where $v=c\frac{E}{H}$. This means that the electrons at the edge
drift with the velocity $v$.

Hamiltonian (energy) of the edge waves is given by
\begin{equation}\label{hamiltonian_edge}
    H=\frac{1}{2}e\int h\rho E dx=\frac{\pi v}{\nu}\int\rho^2dx
\end{equation}
It is easy to quantize this Hamiltonian. Rewriting
\eqref{continuity} and \eqref{hamiltonian_edge} in the momentum
representation and identifying $\rho_k$ with the coordinate
variable one finds that the corresponding canonical momentum is
given by $p_k=2\pi i\rho_{-k}/\nu k$ (zero mode with $k=0$ is
excluded from the theory due to incompressibility of the liquid).
From the commutation relations
$[p_k,\rho_{k^\prime}]=i\delta_{kk^\prime}$ one obtains the
Kac-Moody algebra for the currents
\begin{equation}\label{kac_moody}
    [\rho_k,\rho_{k^\prime}]=\frac{\nu}{2\pi}k\delta_{k+k^\prime}
\end{equation}
This theory provides complete description of low lying excitations
of the Laughlin state.\cite{wen}

Now we want to employ this theory to QSH systems. The easiest way
to understand QSHE is to consider the case when spin $\hat{s}_z$
is conserved.\cite{km1} Then one can define two sectors with spin
up and down. Such decomposition is possible even when spin is not
conserved.\cite{prodan} Then one can define the Chern numbers
$n_{\uparrow,\downarrow}$ for the spin up and down sectors in the
usual manner.\cite{tknn,avron} In a time reversal invariant system
the total Chern number $n_{\uparrow}+n_\downarrow=0$ is zero.
However the difference of the two Chern numbers in general is not
zero and one can define the spin Chern number according to
$C_s=\frac{1}{2}(n_{\uparrow}-n_\downarrow)$. Loosely speaking,
$C_s$ determines the number of gapless edge modes in the system.
It was shown that the edges of a QSH system with even $C_s$ are
localized by disorder, while edges of a QSH system with odd $C_s$
are robust against small time reversal invariant
perturbations.\cite{wbz,xu} The case
$n_{\uparrow}=-n_\downarrow=1$ is the simplest and to describe
edge states of such a system one needs two uncoupled
incompressible liquids $\rho_1$ and $\rho_2$ with the edge state
Hamiltonians \eqref{hamiltonian_edge}, which are related by time
reversal symmetry. Edge excitations travel in opposite directions
with the velocity $u=eE/hn$.  Here the index 1 refers to spin up
and the index 2 refers to spin down sectors.

In analogy with the above case, we assume that the low lying
excitations of a three-dimensional topological insulator are the
surface excitations of two uncoupled incompressible quantum
liquids confined by a potential well, with the following
Hamiltonian:
\begin{equation}\label{hamiltonian_ti}
    H=\int\frac{1}{2}\rho_0eE\left[h_1^2(x,y)+h_2^2(x,y)\right]dxdy
\end{equation}
where $\rho_0$ is the density of the electronic liquid,
$\mathbf{x}=(x,y)$, $x$ and $y$ are coordinates along the surface.
Two-dimensional density of the surface waves is
$\rho_i(x,y)=\rho_0h_i(x,y)$, $i=1,2$. Incompressibility of the
liquid means that $\int h_i(x,y)dxdy=0$. It will turn out that
after making some assumptions about the dynamics of the fields
$h_{i\theta}$, this case corresponds to a strong TI with a single
Dirac cone on the surface.

The essence of Luther's approach is to consider a $D$-dimensional
space as a set of one-dimensional
``rays".\cite{luther1,luther2,aratyn,anderson,sommerfield}
Following these ideas, we write the surface state Hamiltonian
\eqref{hamiltonian_ti}, as a sum over tomographic projections
\begin{equation}\label{hamiltonian_tomographic}
    H=\int_\mathcal{R} d\theta \int \frac{1}{2}neE\left[h_{1\theta}^2(\xi)+h_{2\theta}^2(\xi)\right]d\xi
\end{equation}
integrating over the right hemi-circle $\mathcal{R}=\{-\pi/2\leq
\theta\leq\pi/2\}$. Here the surface displacement of the
tomographically projected system $h_{\theta}(\xi)$ is defined as
\begin{equation}\label{tomographic_projection}
    h_{i\theta}(\xi)=\int_0^\infty
    \left(\frac{k\rho_0}{2\pi^3n}\right)^{1/2}dk\int \cos
    k(\xi-\xi^\prime)h_i(x^\prime,y^\prime)dx^\prime dy^\prime
\end{equation}
$\mathbf{k}=|k|\hat{k}, \hat{k}=(\cos\theta,\sin\theta)$,
$\xi=\hat{k}\cdot\mathbf{x}$,
$\xi^\prime=\hat{k}\cdot\mathbf{x}^\prime$; since $z$ coordinate
remains intact, we can consider the planes $(\xi,z)$ which are
labeled by $\theta$. We call these planes as tomographic planes or
tomographic projections of the initial three-dimensional system.
$n$ is the two-dimensional density of the electronic liquid in the
tomographic plane, which is determined from the consistency of the
two descriptions. Electric field of the confining potential well
is the same in both cases.\footnote{We assume that $E$ is constant
on the surface or varies very slowly as a function of $x$ and
$y$.} It is easily verified that tomographic projections are also
incompressible liquids. Now we make the following assumption that
these tomographic planes are QSH systems. From the topological
band theory we can be sure that this is correct at least for three
values of the parameter $\theta$ (see the discussion at the end of
the paper). The fact that this is correct for all $\theta$ will be
justified below, because this is the only assumption that leads to
the desired result: single Dirac cone on the surface which will be
obtained after fermionization of the model. Then one finds that
excitations have linear spectrum $\omega_k=u|k|$, where $u=eE/hn$.
Thus one can relate the unknown parameter of the theory $n$ to the
parameters of the three-dimensional theory.

The choice of the range of $\theta$ is not unique and this
reflects the fact that splitting of the full Hilbert space induced
by the time reversal operation is not unique. When time reversal
symmetry is preserved such splitting is necessary to obtain a
non-trivial Chern number of a 2D system, because the Chern number
of the whole Hilbert space vanishes (in 3D one needs to consider
certain 2D sections of the Brillouin zone and further split them
using time reversal operation).

One can define Fourier components of $h_{i\theta}(\xi)$:
\begin{equation}\label{}
    \tilde{h}_{i\theta}(k)=\int h_{i\theta}(\xi)e^{-ik\xi}d\xi
\end{equation}
Since by our assumption $\tilde{h}_{i\theta}(k)$ are the edge
modes of a QSH system, they must satisfy the equations of motion
(continuity equations)
$\partial_t\tilde{h}_{1\theta}(k)=iuk\tilde{h}_{1\theta}(k)$ and
$\partial_t\tilde{h}_{2\theta}(k)=-iuk\tilde{h}_{2\theta}(k)$.
From \eqref{tomographic_projection} one also has
$\tilde{h}_{1\theta}(\pm |k|)=\sqrt{\frac{|k|\rho_0}{2\pi
n}}h_1(\mathbf{\pm k})$. This finally leads to
\begin{eqnarray}\label{eqofmotion}
    \partial_t h_1(\mathbf{\pm k})=\pm iu|k|h_1(\mathbf{\pm k})\label{eqofmotion1}\\
    \partial_t h_2(\mathbf{\pm k})=\mp iu|k|h_2(\mathbf{\pm k})\label{eqofmotion2}
\end{eqnarray}
when $\mathbf{k}$ is in the right hemi-circle. In the
second-quantized form the Hamiltonian \eqref{hamiltonian_ti} is
\begin{equation}\label{hamiltonian_bosonized}
    H_B=\sum_{\mathbf{k}\in \mathcal{R}}u|k|(\alpha^\dagger_{\mathbf{k}}\alpha_{\mathbf{k}}+\beta^\dagger_{\mathbf{k}}\beta_{\mathbf{k}})
\end{equation}
where \begin{equation}\label{}
    \nonumber
    [\alpha_{\mathbf{k}},\alpha^\dagger_{\mathbf{k^\prime}}]=
    [\beta_{\mathbf{k}},\beta^\dagger_{\mathbf{k^\prime}}]=\delta_{\mathbf{k},\mathbf{k}^\prime},\quad
     [\alpha_{\mathbf{k}},\beta^\dagger_{\mathbf{k^\prime}}]=0
\end{equation}
e.g, $\alpha_{\mathbf{k}}$ is related to $h_1(\mathbf{k})$ through

$\alpha_{\mathbf{k}}=\sqrt{\rho_0/n|k|}h_1(-\mathbf{k})$,
$\alpha_{\mathbf{k}}^\dagger=\sqrt{\rho_0/n|k|}h_1(\mathbf{k})$
when $\mathbf{k}\in \mathcal{R}$. The Hamiltonian
\eqref{hamiltonian_bosonized} is half of the massless Klein-Gordon
model, exactly what is needed to construct massless two-component
Dirac fermion.\cite{luther1} This is similar to that chiral bosons
in one-dimension are the half of the ordinary bosons. One can
associate the following bosonic fields with this model
\begin{multline}\label{bosonic_field1}
    \phi_1(\theta,\hat{k}\cdot\mathbf{x})=\left(\frac{S}{2\pi^2}\right)^{1/2}\int_{0}^\infty
    e^{-\alpha k/2}dk\\
    \times(\alpha_{\mathbf{k}}e^{ik(\hat{k}\cdot\mathbf{x})}+h.c.)
\end{multline}
\begin{multline}\label{bosonic_field2}
    \phi_2(\theta,\hat{k}\cdot\mathbf{x})=-\left(\frac{S}{2\pi^2}\right)^{1/2}\int_{0}^\infty
    e^{-\alpha k/2}dk\\
    \times(\beta^\dagger_{\mathbf{k}}e^{ik(\hat{k}\cdot\mathbf{x})}+h.c.)
\end{multline}
$\alpha$ is the cutoff which should be taken to zero at the end;
$S$ is the surface area. These fields do not correspond to any
local observables. However, Luther showed that appropriate
functions of these fields do.

Here we review some of the details of the Luther's construction
for completeness.\footnote{Although Luther considered
three-dimensional fermions, it is easy to extend his results to
two dimensions.} Suppose we have a fermionic Hamiltonian
\begin{equation}\label{hamiltonianf}
    H_F=u\sum_{\mathbf{k}}a^\dagger_\mathbf{k}(\vec{k}\cdot\vec{\sigma})a_{\mathbf{k}}
\end{equation}
This is a Hamiltonian of massless Dirac fermions. It can be
diagonalized by a transformation
\begin{equation}\label{transformation}
    U=e^{iS},\quad S=i\sum_{\mathbf{k}}\frac{\pi}{4}a^\dagger_\mathbf{k}\mathbf{V}\cdot
    \vec{\sigma} a_\mathbf{k}
\end{equation}
where $\mathbf{V}=(-\sin\theta,\cos\theta)$. Fermi operators
transform according to
\begin{equation}\label{transformation_fermi}
    a_\mathbf{k}^\prime=e^{-i\pi \mathbf{V}\cdot\vec{\sigma}/4}a_\mathbf{k}
\end{equation}
The diagonal Hamiltonian is
\begin{equation}\label{hamiltonian_diagonal}
    H_F^\prime=\sum_{\mathbf{k}}u|k|a^{\prime\dagger}_\mathbf{k}\hat{\sigma}_za^\prime_{\mathbf{k}}
\end{equation}
The boson representation is given by
\begin{equation}\label{bosonization1}
    \psi(\theta,\hat{k}\cdot\mathbf{x})=e^{-i\pi \mathbf{V}\cdot\vec{\sigma}/4}\psi^\prime(\theta,\hat{k}\cdot\mathbf{x})
\end{equation}
\begin{equation}\label{bosonization2}
    \psi^\prime(\theta,\hat{k}\cdot\mathbf{x})=\frac{1}{2\pi\alpha}\begin{pmatrix}
      \exp[\phi_1(\theta,\hat{k}\cdot\mathbf{x})] \\
      \exp[\phi_2(\theta,\hat{k}\cdot\mathbf{x})] \\
    \end{pmatrix}
\end{equation}
(Klein factors necessary to ensure anticommutation relations are
omitted for simplicity; for details see
[\onlinecite{luther1,luther2}]). The usual Fermi fields are given
by
\begin{equation}\label{fermi_operator}
    \psi(\mathbf{x})=\int_\mathcal{R}d\theta\psi(\theta,\hat{k}\cdot\mathbf{x})
\end{equation}
The representation
\eqref{bosonic_field1},\eqref{bosonic_field2},(\ref{bosonization1}-\ref{fermi_operator})
is constructed in such a way that the correlation functions of
free fermionic fields are correctly reproduced. Note a crucial
point that in the definition \eqref{fermi_operator} the
integration is only over half of the whole circle. Thus we see
that the Hamiltonian \eqref{hamiltonian_ti} together with the
equations of motions \eqref{eqofmotion1} and \eqref{eqofmotion2}
(or alternatively the Hamiltonian \eqref{hamiltonian_tomographic}
where $h_{i\theta}(\xi)$ are edge states of a QSH insulator) is
equivalent to massless Dirac Hamiltonian. This corresponds to
strong TI with a single Dirac cone on the surface. These
calculations confirm our initial intuition.

There is only one $Z_2$ invariant in two-dimensions.\cite{km} It
can be defined as $C_s$ mod $2$.\cite{sheng} In three-dimensions
there are four $Z_2$ invariants.\cite{fkm,moore,roy} Three of them
are equivalent to invariants of two-dimensional topological
insulators and are defined as invariants of some sections of the
Brillouin zone. The fourth topological invariant $\nu_0$ is purely
three-dimensional. TIs with $\nu_0=0$ are called weak and TIs with
$\nu_0=1$ are called strong. This invariant determines the number
of Kramers degenerate Dirac points enclosed by the Fermi surface.

In any time reversal invariant system with spin orbit interactions
there are two-dimensional Dirac points in the surface spectrum.
Therefore the above considerations should be clarified. Single
Dirac fermion on the surface already means that the insulator is a
strong topological insulator. We will show this directly and in
parallel discuss the relation of our picture to the conventional
theory of topological insulators. To make connection of this
picture with the band topology, we use the simple argument for
counting the topological invariants due to Roy\cite{roy2}, which
is quoted below. For us the important aspect of this work is how
the $Z_2$ invariants of certain planes in the Brillouin zone, such
as $p_x=p_y$, can be computed.

Let represent the Brillouin zone by a cube $\{-\pi\leq
p_x,p_y,p_z\leq\pi\}$ and the $Z_2$ invariants associated with the
planes $p_x=0$, $p_x=\pi$, $p_y=0$, $p_y=\pi$ by
$\nu_1,\tilde{\nu}_1,\nu_2,\tilde{\nu}_2$ respectively. Then the
$Z_2$ invariant of the plane $p_x=p_y$ equals
$\nu_1+\tilde{\nu_2}$ and the $Z_2$ invariant distinguishing
strong topological insulators from weak topological insulators
equals $\nu_0=\nu_1-\tilde{\nu}_1=\nu_2-\tilde{\nu}_2$. Any 3D
topological insulator with time reversal invariance can be
characterized by four invariants, which may be chosen to be
$\nu_1,\nu_2,\nu_3$ and $\nu_0$.

Above we assumed that every tomographic plane supports QSHE. In
fact, it is sufficient to consider only three planes (then for the
rest this would be satisfied automatically). Let these planes be
$p_x=\pi$, $p_y=\pi$, $p_x=p_y$. We will consider for concreteness
the Dirac point $(\pi,\pi,0)$ and the surface states on the
$(x,y)$ plane having small 2D momentum $\mathbf{k}$ around
$p_x=\pi$,$p_y=\pi$ (direction of $\mathbf{k}$ is given by the
angle $\theta$). In this case $\theta=0$ corresponds to the plane
$p_y=\pi$, $\theta=\pi/2$ to the plane $p_x=\pi$, and
$\theta=\pi/4$ to the plane $p_x=p_y$. Suppose that $Z_2$
invariant of each of this planes is odd, thus corresponding to
non-trivial insulator. Then we have
$\tilde{\nu}_1=\tilde{\nu}_2=\nu_1+\tilde{\nu}_2=1$, which gives
$\nu_0=1$. This corresponds to strong topological insulating
phase. If there is no such a point in the Brillouin zone, for
which all three $Z_2$ invariants are 1, then the insulator is in
the weak topological insulating phase with $\nu_0=0$. (see also
Fig.3 in the Ref. [\onlinecite{inversion_symmetry}])

In general, it isn't possible to define a $Z_2$ invariant of a
plane with arbitrary $\theta$ using topological band theory. It
seems that this is possible only for planes such as
$\tan\theta=m/n$, where $m$ and $n$ are two co-prime integers. In
the case we have considered, all such invariants are equal to 1.

Recently, bosonization approach was applied to topological
insulators also in the work [\onlinecite{moore2}], but in a
different context. In passing we also note that surface
excitations of certain 3D topological superconductors\cite{topsc}
can be viewed, in every radial direction, as edge states of a 2D
topological superconductor in the same class. It would be
interesting to explore this case as well.

In summary, we have shown that low lying excitations of a strong
TI with a single Dirac cone on the surface are the surface
deformations of a droplet of incompressible quantum liquid. These
excitations have very unusual form, however they have a simple
meaning when one considers tomographic projections of this liquid
(as defined in the text): they are two chiral waves propagating in
opposite directions. Thus, surface excitations of a strong TI with
a single Dirac cone on the surface are the sum of QSH edge states.
However, this is not true for the entire topological insulator,
i.e., a strong TI can not be presented as a sum of 2D TIs. This
paper can be considered as an another physical illustration of the
fact, that 3D topological insulators can be characterized by 2D
invariants.

I would like to acknowledge P.I. Arseev, A.G. Semenov and
especially S.M. Apenko and V.V. Losyakov for numerous useful
discussions and comments on the earlier versions of this
manuscript and Yu.E. Lozovik for drawing my attention to
topological insulators.

\end{document}